# Time, Space and Structure in Ancient India

Subhash Kak

**Introduction**

This essay uses Vedic and classical sources to present a synoptic vision of the universe in ancient India and shows its continuity across different periods. This vision was based on an assumed equivalence of the outer and the inner cosmoses and it is embodied in architecture, music, and art. It provides an archaeoastronomical window on Indian monumental architecture.

The ancient world did not have a split between the sacred and the temporal. The temple served as the place where time-bound ritual was conducted and keeping time was one of its functions. The English word *temple* is derived from the Latin *templum,* which is sacred space, facing west, that was marked out by the augurs. In the east-west orientation of the axis of the temple that is strictly true only on the two equinoxes is the acknowledgement of concern with time and the seasons.

In India, the temple is likewise associated with the east-west axis and we can trace its origins to priests who maintained different day counts with respect to the solstices and the equinoxes. Specific days were marked with ritual observances [1] that were done at different times of the day. Some ritual included construction of altars that coded knowledge related to the motions of the sun and the moon and supposed correspondences with the inner cosmos.

The Agnicayana altar, the centre of the great ritual of the Vedic times that forms a major portion of the narrative of the Yajurveda, is seen as the prototype of the temple and of the Indian tradition of architecture (Vāstu). The altar is first built of 1,000 bricks in five layers (that symbolically represent the five divisions of the year, the five physical elements, as well as five senses) to specific designs [2]. The Agnicayana ritual is based upon the Vedic division of the universe into three parts of earth, atmosphere, and sky, which are assigned numbers 21, 78, and 261, respectively; these numbers add up to 360, which is symbolic representation of the year. These triples are seen in all reality, and they enlarge to five elements and five senses in further emanation.

In the ritual at home, the householder employed three altars that are circular (earth), half-moon (atmosphere), and square (sky) at his home (Figure 1), which are like the head, the heart, and the body of the Cosmic Man (Purusha). In the Agnicayana ritual, the atmosphere and the sky altars are built afresh in a great ceremony to the east. The numerical mapping is maintained by placement of 21 pebbles around the earth altar, sets of 13 pebbles around each of 6 intermediate (13×6=78) altars, and 261 pebbles around the great new sky altar called the Uttara-vedi.

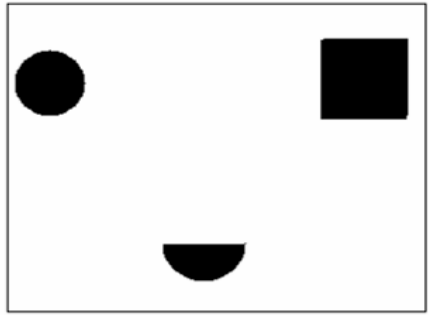

**Figure 1.** The three altars of the Vedic house: circular (earth, body), half-moon (atmosphere, prāna), square (sky, consciousness)

The Uttara-vedi is the precursor to the temple structure [3]. It also symbolizes the patron in whose name the ritual is being performed [4].

The underlying bases of the Vedic representation and ceremony are the notions of *bandhu-* (equivalence or binding between the outer and the inner), *yajña* (transformation), and *paroksha* (paradox). As mentioned before, the five layers of the altar represent the physical world, and the number of consecrated bricks in the five layers is related to numbers derived from the year count of 360. To represent two more layers of reality beyond the purely objective, a sixth layer of bricks that includes a hollow brick with an image of the golden Purusha inside is made, some gold chips scattered and the fire placed, which constitutes the seventh layer (ŚB 10.1.3.7). The two layers beyond denote completion, for seven was taken as a measure of the whole. The symbolic meaning of this is that the ceremonies of the great altar subsume all ritual [5].

**Recursion**

The central idea of this scheme is that of *recursion,* or repetition in scale and time. The universe is taken to be mapped into the individual; it is also symbolically represented in the creative arts, as in music, dance, sculpture, and sacred architecture [4],[6],[7]. In literature, we see recursion in the *story within story* genre that is to be found in the Vedic hymns, the Epics, the Yoga Vasishtha, and the Puranic texts.

Not only is the temple a symbolic representation of the cosmos, the Rigveda itself was planned as a five-layered altar by stacking up the 10 books in pairs, two books to a layer [2], as shown in Figure 2. These hymn numbers have several symmetries, such as pairs of hymn numbers differing by 12, 17 and 29, and the numbers have an astronomical basis that is described at length in *The Astronomical Code of the Rgveda* [2].

| 191 | 114 |
|---|---|
| Book 10 | Book 9 |
| 104 | 92 |
| Book 7 | Book 8 |
| 87 | 75 |
| Book 5 | Book 6 |
| 62 | 58 |
| Book 3 | Book 4 |
| 43 | 191 |
| Book 2 | Book 1 |

**Figure 2.** The Rigveda as an altar [2]

The separations diagonally across the two columns are 29 each for Book 4 to Book 5 and Book 6 to Book 7, and they are 17 each for the second column for Book 4 to Book 6 and Book 6 to Book 8. Books 5 and 7 in the first column are also separated by 17; Books 5 and 6 as well as 7 and 8 are separated by 12; Books 5 and 7 also add up to the total for either Book 1 or Book 10. Another regularity is that the middle three layers are indexed by order from left to right whereas the bottom and the top layers are in the opposite sequence.

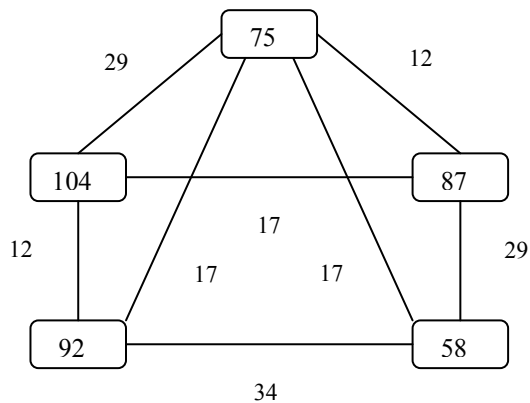

**Figure 3.** The Rigveda books 4-8 as a graph

The relationship between the hymn numbers for the books 4-8 looks particularly striking when viewed through the graph of Figure 3, where the numbers on the edges represent the difference between the corresponding node numbers.

The total number of hymns, 1017, is $3 \times 339$, where 339 is the number of sun-disks from sunrise to sunset on a typical day (or similarly of moon-disks). The origin of the number 339 becomes evident if the sun were to be 108 times its diameter away from the earth (which is astronomically true [3]) as $108 \times \pi \approx 339$. Note also that the sum of the sky and the atmosphere numbers is $261+78 = 339$, indicating that the choice of the three numbers must have had an observational basis.

With the temple viewed as a map of the universe, the main altar corresponds to the sun and the door at the west corresponds to the earth. The origin in India of the measure of 108 as the distance in sun-diameters from the earth to the sun may be the discovery that a pole of a certain height removed to a distance of 108 times its height has the same angular size as the sun or the moon. It led to a conceptual framework for the sun-earth-moon system that became a part of Indian cosmology. It was further assumed that beyond our system existed other worlds.

Owing to assumed recursion, the number 108 characterizing the cosmos should also be associated with the gods and the activities of humans. Not surprisingly, this number shows up as the number of beads in the Indian rosary (telling of beads is to make a symbolic journey across the worlds), the number of dance movements of the Nātya Śāstra, the names of the God and the Goddess, the number of pilgrimages, the number of spiritual masters, and so on. There are also 108 divisions of the zodiac and 108 rhythmical patterns (tālas) of music. With the human body described by a measure of 108, the weak points of the body are counted in the Āyurveda system to be 107.

Likewise, each of the 27 nakshatra divisions of the zodiac is further divided into 27 upa-nakshatras (ŚB 10.5.4.5). The time measures are defined in a sequence with multiples of 30.

**Ritual and Plan of the Temple**

We now briefly summarize our work [3] on the axis and the perimeter of the sacred ground as available to us in the Śatapatha Brāhmana, which validates the astronomical interpretation of 108.

The Agnicayana ritual, as sacred theatre [8], was performed in a special area where the three fires of the yajamāna are established in the west in an area called Prācīnavamsha, "Old Hall" whose dimensions are in the canonical ratio of 1:2.

The Prācīnavamsha has dimensions of 20×10. Three steps from it to the east (ŚB 3.5.1.1) is the Mahāvedi, which is an isosceles trapezoid of spine 36 and the two sides of 30 and 24 units, representing the ratio of 5/4. To see the significance of the plan, we now draw the sacred ground within a rectangular area (Figure 4).

For accord with the measures which are multiples of 6, the left area is increased by an additional one step to the west to become 24×30 as in Figure 4, which is described as an appropriate proportion for a house in later texts such as Varāhamihira's Brihat Samhitā (53.4) [9]. The Prācīnavamsha's contribution to the perimeter is 24+30+24=78, which is the atmosphere number that we have come across before. This is also in accord with the notion that the Prācīnavamsha is tripled in size in the completion of the Mahāvedi, going from 10×20 to 30×60. The distance to the high altar on the extreme right (with dimensions of 6×6) from the axis at the left is 54 units. The high altar is where the main ritual is performed and, symbolically, it represents the sun. The separation of the high altar from the doorway to the left is representative of the distance to the sun and the perimeter is representative of the circuit of the sun.

Thus the basic temple plan contains two significant numbers, 180 and 54, which, when doubled, correspond to astronomical knowledge related to the 360 days of the year (attested in the Rigveda) and the number 108 (distance to the sun in sun-diameters).

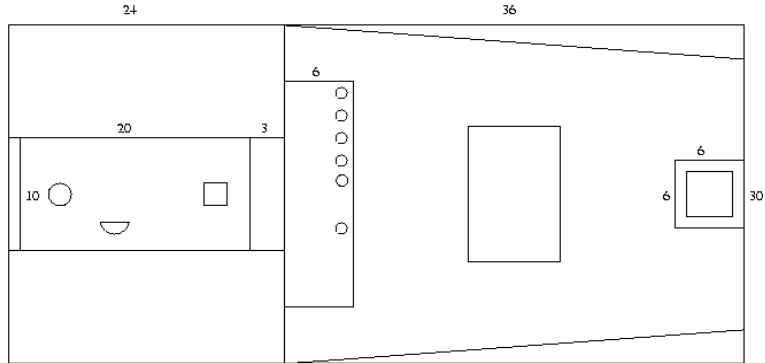

**Figure 4:** The temple plan: The perimeter is 180 units and the axis is 54 units to the high altar

The proportion of 1:2 for the altar ground is attested in later texts such as the Brihat Samhitā [9] and Śilpa Prakāsha [10]. The altar ground is also associated with ratios of 5/4, 3/2, and so on, which emerge out of the interplay of numbers such as 30 days of the month, 180 days of the half-year, 12 months of the year, and so on. It must be remembered that astronomical numbers are also associated with the ratios of the seven notes of the octave [11].

The temple itself, in its three-dimensional form, codes several rhythms of the cosmos and specific alignments related to the geography of the place and the presumed linkages of the deity and the patron. The architecture may also incorporate themes related to royal power if it was built at the behest of a king.

As mentioned before, in the Vedic system the earth was assigned the form of circle and the sky the form of square. The main altar in the temple has a square form because it represents the sky, with the four corners standing for the four cardinal directions or the set of two equinoxes and two solstices [6].

Buildings to be used by humans are to depart from the perfect, square form. The breaking of the symmetry is to connote energy and change. According to the Brihat Samhitā (chapter 53) [9], the length of a king's palace is greater than the width by a quarter (1 + ¼ = 5/4), or in other words the proportions are 5:4. Likewise, the length of the general's palace exceeds the width by a sixth (1 + 1/6 = 7/6), or has proportions 7:6; the proportions for the houses of ministers and princes are 9:8 and 4:3, respectively. For lesser officials as well as citizens, other proportions are prescribed.

**Recursion and the Śri Cakra**

Yantras have been used in India to represent esoteric knowledge, and the basis of this esoteric knowledge is cosmology. The Śri Cakra or Śri Yantra (Figure 5) is the tripartite division of the universe into earth, atmosphere, and the sun mirrored in the individual by the body, the breath, and the inner lamp of consciousness. It may also be seen as the three main divisions of the body into the head, the chest, and the lower trunk. Its basic form is that of three triangles, two pointing downwards and one upwards. Within each triangle are two other triangles, of alternating opposing polarity in terms of the direction of pointing that represents male and female principles. All together, this adds up to 9 interpenetrating triangles (5 downward pointing Śakti principle and 4 upward pointing Śiva principle), which through their overlaps constitute a total of 43 small triangles. Right through the middle of this is the dot, the *bindu,* who is Śiva, the witness, or consciousness.

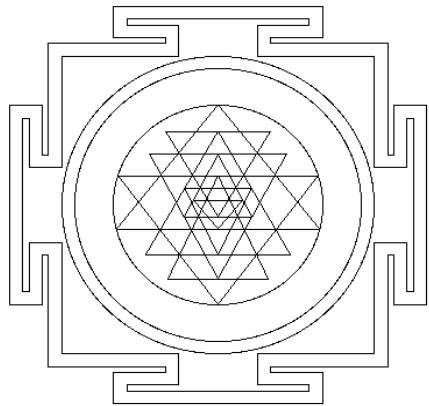

**Figure 5.** The Śri Cakra

These triangles formed by the intersection of the nine triangles are surrounded by a circle of 8 petals that, in turn, is surrounded by a 16-petalled circle (petals not shown in the figure above). At the outermost are lines, which are called the *bhūpura* (the city of the earth, or the body). The yantra is also categorized into 9 coverings (āvaranas), where the *bhūpura* is the outer covering. These 9 coverings have 108 presiding Goddesses. In the Sri Cakra pūja they are systematically worshipped one by one with their names and mantras.

The presiding deity of Sri Cakra is Lalitā Tripura Sundarī, who is the Goddess in her playful aspect as the transcendent beauty of the three worlds. These three worlds are the three gunas of *sattva*, *rajas* and *tamas*; or sun (heavens), moon (atmosphere), and fire (earth); will (*icchā*), knowledge (*jñāna*) and action (*kriyā*); intellect, feelings, and sensation; subject (*mātā*), instrument (*māna*), and object (*meya*) of all things; or waking (*jāgrat*), dreaming (*svapna*) and dreamless sleep (*susupti*) states. Her five triangles represent the *pañca bhūtas* (five elements). She holds five flowery arrows, noose, goad and bow. The noose is attachment, the goad is revulsion, the bow is the mind and the flowery arrows are the five sense objects.

The Śrī Cakra ritual infuses the design of the yantra with chant, representing the union of space and sound. Its closed, concentric circuits correspond to the nine planes of consciousness of the spiritual seeker. Each plane is a stage on the ascent of one's being toward the inner self.

**Cosmic Vision in the Sindhu-Sarasvati World**

Does a cosmic vision characterize the Sindhu-Sarasvati (Harappan) civilization also? For this we need evidence of astronomical alignments, use of specific proportions, and that of the important concept of recursion.

Archaeologists agree that there is continuity in religion, art, and culture [12]-[16] between the Sindhu-Sarasvati period of the third millennium BC and the later historical period; further supporting evidence is summarized in [17]-[21]. Speaking now of a quantitative measure, we saw that the number 108 is central to Indian cosmology; it is also an essential component of the Indian system of length units going back to the Harappan period. Although this in itself does not establish that the Harappan cosmology is identical to the Vedic (for the use of the number 108 could be a coincidence), but the importance given to astronomy in the Harappan region and the other cultural correspondences suggests that it was so. The evidence from Dholavira, as described later in this section, suggests that the idea of recursion was part of the worldview of the Harappans.

Maula [22] presents evidence on the use of great calendar stones, in the shape of ring, which served to mark the beginning and end of the solar year in Mohenjo-Daro indicating that astronomy had moved beyond marking the motions of moon. For comparison, note that the astronomical basis of the Vedic ritual was the reconciliation of the lunar and solar years.

Wanzke [23] argues that Mohenjo-Daro and other sites show slight divergence of $1°$ to $2°$ clockwise of the axes from the cardinal directions. He suggests that this might have been due to the orientation of Procyon and Aldebaran, two bright stars that were prominent in the direction of the setting sun during 2500 B.C. to 1500 B.C., indicating once again astronomical reasoning of the kind to be found in the design of Hindu temples [6].

It is significant that yantric buildings have been discovered in North Afghanistan that belong to a period that corresponds to the late stage of the Sindhu-Sarasvati civilization [4] providing architectural evidence in support of the idea of recursion at this time. Although these building are a part of the Bactria-Margiana Archaeological Complex (BMAC), their affinity with ideas that are also present in the Sindhu-Sarasvati system shows that these ideas were widely spread.

Some Harappan seals are generally accepted to have an astronomical basis, and the beginnings of the Vedic nakshatra system are seen in the third millennium BC. There is also continuity in the system of weights and lengths between the Harappan period and the later historic period [24].

Danino [25],[26] provides an analysis of the unit of length at Dholavira that is in accord with unit from the historical period. He shows that the unit that best fits the Dholavira dimensions is 190.4 cm, which when divided by 108 gives the Dholavira angula of 1.763 cm. The subunit of angula is confirmed when one considers that the bricks in Sindhu-Sarasvati follow ratios of 1:2:4 with the dominating size being 7 × 14 × 28 cm ([15], page 57). These dimensions can be elegantly expressed as 4 × 8 × 16 *angulas*, with the unit of *angula* taken as 1.763 cm.

It is significant that the ivory scale at Lothal has 27 graduations in 46 mm, or each graduation is 1.76 mm.

Figure 6 is the general plan of Dholavira, which consists of three "towns" in accord with Vedic ideas [16],[20]. Figure 7, taken from [26], presents the proportions for the three towns of Dholavira and shows how they are related. The Dholavira system of units, due to Danino, is clearly shown in Figure 8. It was found by using units that led to integer counts for lengths with a small margin of error.

The feature of recursion in the three towns, or repeating ratios at different scales, is significant because we have already noted it in the Vedic world view. Specifically, the design is characterized by the nesting proportion of 9:4 across the lower and the middle towns and the castle. The proportions of 5/4, 7/6, and 5/4 for the lower town, the middle town, and the castle may reflect the measures related to the royal city, the commander's quarter, and the king's quarter, respectively. We don't know if the choice of the other proportions in Figure 8 had a definite logic behind it.

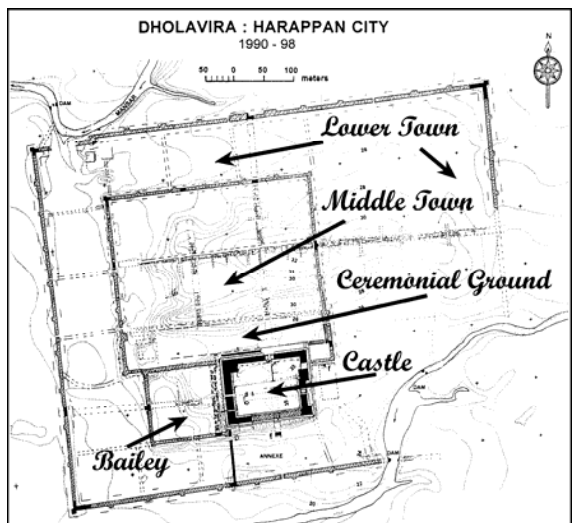

**Figure 6.** Map of Dholavira (Bisht [16])

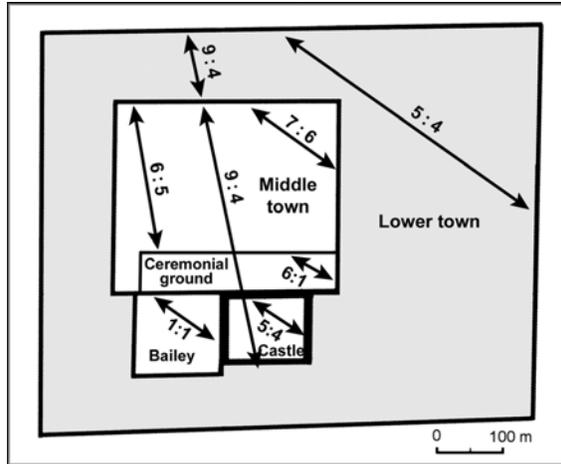
**Figure 7.** Proportions from Dholavira (Danino [25])

Danino's analysis [26] of the Dholavira length unit D shows that it corresponds to the Arthaśāstra 2.20.19 measure of dhanus that equals 108 angulas. This scale is confirmed by a terracotta scale from Kalibangan and the ivory scale found in Lothal. The Kalibangan scale [27],[28] corresponds to units of 17.5 cm, which is substantially the same as the Lothal scale and the small discrepancy may be a consequence of shrinkage upon firing.

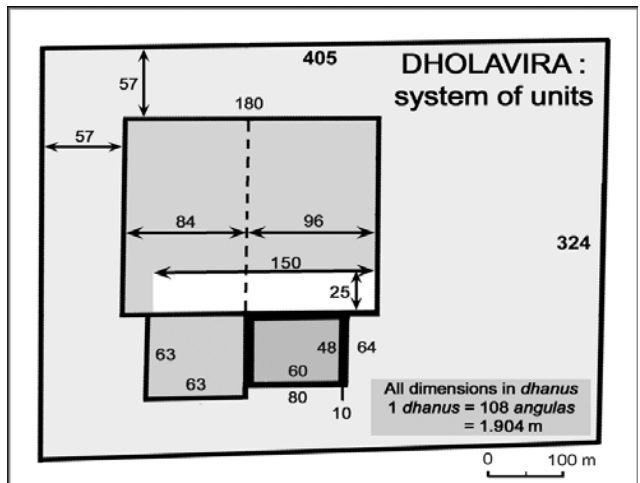
**Figure 8.** The dimensions of Dholavira in terms of dhanus (Danino)

Pant and Funo [29] find continuity between the grid and modular measures in the town planning of Mohenjodaro and Kathmandu Valley. They find the measure of 19.2 meters as a unit in quarter-blocks of the town; this is nearly the same as the unit characteristic of the dimensions of Dholavira. It shows that the traditional architects and town planners have continued the use of the same units over this long time span.

**Historic Period**

Balasubramaniam has found new evidence supporting the continuing use in the historical period of the 108-sub-unit based measurement scheme in a variety of monumental architecture. Recently, he found [30] confirmation of the use of the unit of dhanus (D) in the Delhi iron pillar (Figure 9), and he showed that its dimensions are elegantly related to D.

    Underground base:       4/5 D
    Height of main pillar:  12/5 D
    Capital:                4/5 D
    Diameter at foundation: 3/9 D
    Diameter at ground:     2/9 D
    Diameter at the top:    1/9 D

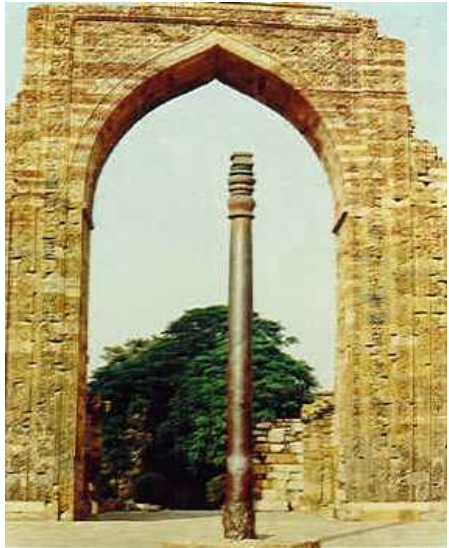
**Figure 9.** The Delhi Iron Pillar

The division of D into 108 parts is established by the fact that the diameter at the top, at the ground and the base are multiples of 12 angulas.

**Recursion in the Deogarh temple**

The Dashavatara Vishnu temple at Deogarh [31] is one of the oldest surviving temples from India, and it is dated to about 500 A.D. (Figures 10 and 11). It is significant that this temple is modular with the basic shape repeated to different scales. Lubotsky [32] has identified it with the Sarvatobhadra temple described in the Vishnudharmottara Purana. The Brihat Samhita 53.31 calls a building with a verandah running all around as "sarvatobhadra" [9].

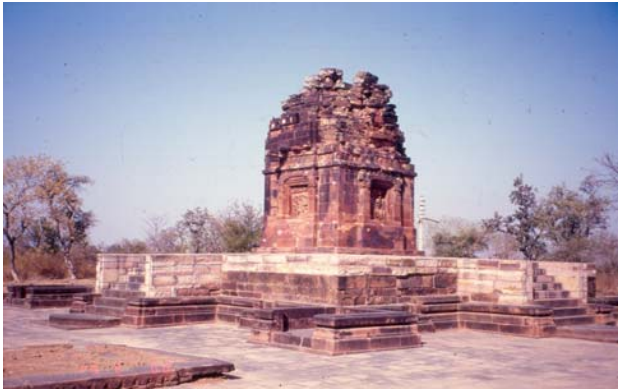
**Figure 10.** The Deogarh temple

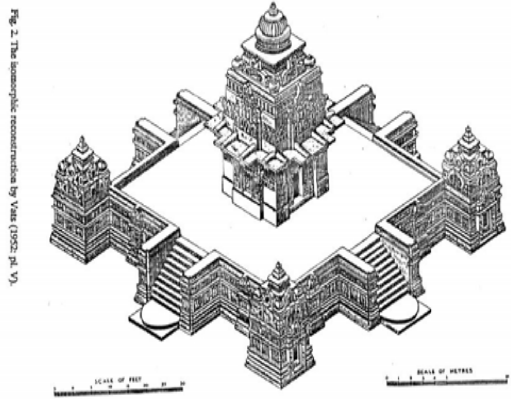
**Figure 11.** Reconstruction of Deogarh temple by Vats

Lubotsky identifies the overall system of the temple with the Pāñcarātra doctrine and shows that it incorporates the four emanations of Viśnu. Meister [33] shows how this temple creates a model that makes symbolic substitutions in stone of complicated architectural forms. At the overarching level, the central structure is repeated to a lesser scale in the four directions. This idea is to be found in the architecture of other temples as well [5].

The idea of recursion underpins Indian arts [34], as it does cosmology and medicine [35] in the historic period.

**Conclusions**

The aim of this essay is to show continuity from the Sindhu-Sarasvati phase of the third millennium BC to the historic period in the ideas of tripartite division and of recursion. Dholavira presents a general plan of the city in terms of its three divisions where the proportions are defined recursively and in ratios that are also to be found in the Vedic temple. It is likely that the notion of the equivalence of the microcosm and the macrocosm, which is at the heart of the tripartite division and recursion in the Vedic period, had currency in the Sindhu-Sarasvati phase.

More specifically, we find the use of the same unit of length (dhanus) with its 108 parts both in the Sindhu-Sarasvati and the historic periods. The use of a scale with 108 divisions is significant because it reflects fundamental ideas related to the nature of the cosmos. This parallel, by itself, could be a coincidence, but when viewed together with the continuity in religion, art, and architecture, it demonstrates a common vision of the universe.

**Acknowledgement**. I would like to thank R. Balasubramaniam and M. Danino for comments on earlier versions of this paper.